# Bringing personalized learning into computer-aided question generation

Yi-Ting Huang, Meng Chang Chen, and Yeali S. Sun

**Abstract**—This paper proposes a novel and statistical method of ability estimation based on acquisition distribution for a personalized computer-aided question generation. This method captures the learning outcomes over time and provides a flexible measurement based on the acquisition distributions instead of pre-calibration. Compared to the previous studies, the proposed method is robust, especially when a student's ability is unknown. The results from the empirical data show that the estimated abilities match the actual abilities of learners, and the pre-test and post-test of the experimental group show significant improvement. These results suggest that this method can serves as the ability estimation for a personalized computer-aided testing environment.

——————— ◆ ———————

## 1 INTRODUCTION

RECENT theories on learning have focused increasing attention on understanding and measuring student ability. There is now general consensus over Vygotsky's [1] observation that a learner's ability in the Zone of Proximal Development (ZPD)—the difference between a learner's actual ability and his or her potential development—can progress well with external help. Instructional scaffolding [2], closely related to the concept of ZPD, suggests that appropriate support during the learning process helps learners achieve their learning goals. Effective instructional support requires identifying students' prior knowledge, tailoring assistance to meet their initial needs, and then removing this aid when they acquire sufficient knowledge.

Nowadays, ability estimation based on Item Response Theory (IRT) [3],[4],[5] offer extensive applications of various domains in e-learning systems. For example, Chen et al. [6] considered a learner's ability for recommending personalized learning paths in a Web-based programming learning system, while Chen and Chung [7] analyzed students' understanding by suggesting English vocabulary on mobile devices. Similarly, within Computerized Adaptive Testing (CAT), Barla et al. [8] calculated an examinee's ability to select suitable questions. All of these studies demonstrated their systems could better adapt to students' educational needs and improved student performance.

Educational data mining is an emerging and rapidly growing field, which developing computational approach for analyzing large-scale data in order to understand how students learn and give them better supports. Data collecting from student responses of their click choices, their texts or symbolic inputs over practice attempts has been investigated in student modeling. For instance, research on and expansion of the Bayesian Knowledge Tracing models [9] has been widely used to identify whether a student has master a specific skill in Intelligent Tutoring Systems (ITS), such as [10], [11], [12], [13] and [14].

Although the purposes of these applications are similar, they need human effort involved. Klinkenberg et al. [15] noted that the Item Response Theory was more suitable for measurement only, one of reasons being that the parameters of items had to be pre-calibrated in advance before items were used in a test. Generally, during the item calibration, an item should be taken by a large number of people, ideally between 200 to 1000 people, in order to estimate reliable parameters for the items [16],[17]. In addition, existing models used for the field of Educational Data Mining, they need experts involving in systems, i.e. experts identified difficulty factors for knowledge components. These procedures are very costly and time-consuming, and also impractical for learning with online resources.

The number of new documents and language material uploaded online is growing at a seemingly exponential pace. With the rapid development of the field of e-learning, learners are getting used to acquiring knowledge, practicing and assessing themselves online. One of the more promptly advancing subfields is computer-aided question generation. It automatically generates questions and learners can actively exercise when a learning material is given. There is a multitude of studies now available for designing different question types, such as multiple–choice questions [18],[19],[20],[21],[22],[23], [24],[25], cloze tests [26],[27],[28], and TOEFL synonym questions [29]. However, it is difficulty to directly apply the previous studies, i.e. Item Response Theory or Bayesian Knowledge Tracing, to estimate the proficiency level of a learner, because they need item parameters in advanced.

In this paper, we present an alternative method of es-

---

- *Y.T. Huang is with the Institute of Information Science, Academia Sinica, Taipei, Taiwan. E-mail: ythuang@iis.sinica.edu.tw.*
- *Meng Chang Chen is with the Institute of Information Science, Academia Sinica, Taipei, Taiwan. E-mail: mcc@iis.sinica.edu.tw.*
- *Yali S. Sun is with the the Department of Information Management, National Taiwan University, Taipei, Taiwan. E-mail: sunny@ntu.edu.tw.*

timating ability based on acquisition distribution. It is designed for a personalized computer-aided question generation, called AutoQuiz. The term "personalized" in this paper refers to the adjustments to students' needs by matching the difficulty level of questions to their knowledge level. This method draws a connection between students' abilities and the acquisition age distributions. For example, if a student is estimated as a certain grade level (i.e. grade level six), this method assigns this grade because the student has acquired a certain percentage of the knowledge in a population, when the student correctly answers a certain percentage of items in a test. Moreover, this method captures the succession of learning over time by employing Exponential Moving Average (EMA) to combine the historical data with the current ability. We also conduct a simulation study to investigate the property of the proposed approach and an empirical study to evaluate practical performance. The research questions addressed in this study are:

1. What are the characteristics of the proposed ability estimation based on the acquisition distribution?
2. How is the performance of the proposed ability estimation compared with the other ability estimations?
3. How is the performance of the proposed ability estimation when students are continuously learning?
4. What is the performance of the proposed ability estimation with the empirical data in a computer-aided question generation testing environment?
5. Does the appropriate instructional scaffolding help students advance their learning progress when their abilities are effectively identified by the proposed ability estimation?

## 2 RELATED WORK

### 2.1 Test theories and their applications

Classical test theory [30], also called as true-score theory, assumes that an examinee's ability can be seen as a true test score. An observed score from a test is composed of a true score and an error. A typical item analysis relies on evaluating item difficulty and discrimination of a test, as well as the reliability of test scores based on a random sample of examinees from the population. It has been developed for several decades and is quite straightforward to apply. However, its major limitation is that the parameters are dependent on examinees and tests. That is, examinees will have higher scores on easier tests, but lower scores on difficult ones. It is hard to get consistent parameters over a number of tests.

Unlike Classical Test Theory, an ability parameter and item parameters of Item Response Theory are invariant. Item Response Theory [3],[4],[5] is a modern theory of testing that examines the relationship between an examinee's response and an item related to an ability measured by items in a test. Three well-known ability estimations proposed by Item Response Theory are maximum likelihood estimation (MLE), maximum a posteriori (MAP) and expected a posteriori (EAP). The procedure of MLE, an iterative process, is to find the maximum likelihood of a response to each item for an examinee. MAP [31] and EAP [32], which are variants of Bayes Modal Estimation (BME), incorporate prior information into the likelihood function. Prior distributions can protect against outliers that may have negative influence on ability estimation. For example, Barla et al. [8] employed EAP to score each examinee's ability for each test. Recently, Lee [33] proposed an alternative computational approach in which a Gaussian fitting to the posterior distribution of the estimated ability could more efficiently approximate that determined by the conventional BME approach. Wang, Berger and Burdick [34] proposed dynamic item response models, which incorporating time, learning growth and the nature of the daily and test random effect, in order to deal with the violation of the local dependence assumption in CAT.

Klinkenberg et al. [15] introduced a new ability estimation based on Elo's [35] rating system and an explicit scoring rule. Elo's rating system was developed for chess competitions and used to estimate the relative ability of a player. With this method, pre-calibration was no longer required, and the ability parameter was updated depending on the weighted difference between the response and the expected response. This method was employed in a Web-based monitoring system, called a computerized adaptive practice (CAP) system, and designed for monitoring arithmetic in primary education.

### 2.2 Student modeling in educational data mining

An emerging field of educational data mining concerned with the application of data mining, machine learning, and statistics to data from education technologies, such as Intelligent Tutoring Systems, and Massive Open Online Courses (MOOC). The topic of student modeling, assessing students' knowledge level, mental models, preferences and needs, has received considerable attention. Two well-known statistical models for estimating students' skill proficiency are Knowledge Tracing and logistic regression model family. Knowledge Tracing [9] can be described as a Hidden Markov Model, models the student's knowledge as latent variables and observes the responses to questions as the acquisition of knowledge in order to capture the dynamic of knowledge probabilistically. On the other hand, Logistic regression models extend Item Response Theory and include additional of factors in the Linear Logistic Test Model. For example, Additive Factors Model (AFM) is based on logistic regression and consider three parameters: students, skills and learning rates [36],[37]; Performance Factors Model (PFM) [38] extends AFM and takes the correctness of individual responses in the previous stage into consideration; Instructional Factors Model (IFM) adds a parameter to represent instructional interventions [39].

### 2.3 Age of acquisition and its potential possibility

The basic idea of acquisition distribution originates from the age of acquisition, the age at which a word, a concept, even specific knowledge is acquired. For instance, people learn some words such as "dog" and "cat" before others such as "calculus" and "statistics". Numerous studies in psychology and cognitive science have shown the influence on the process of brain, such as object recognition [40], object naming [41], [42], [43], [44], and language learning [45], [46], [47], [48].

Today, this concept can be realized with advanced technology, Information Retrieval [49] [50] and Natural Language Processing [51], which counts word frequency and calculates the probability of which a word is acquired when given a group of documents. With a large enough resource, such as an extensive collection of all learning materials which people read and learn, knowledge acquisition age distributions can be computed and implemented. For example, based on textbooks authored specifically for students in grade level six, questions can be generated based on concepts in these textbooks that were correctly answered by a student, and from this, the student can be said to either have or lack the skills at the grade level six. This implies that learning materials, such as textbook, are written with intent to represent what learners at a certain grade level learn and acquire. Two related work to this concept are a readability prediction [52], which mapped a document to a numerical value corresponding to a grade level based on the distribution of acquisition age, and a word difficulty estimation [53], which modeled language acquisition with Latent Semantic Analysis to compute the degree of knowledge of words at different learning stages.

## 3 METHOD

### 3.1 Testbed application

The measurement approach proposed in this study is implemented on a Web-based learning system developed by the AutoQuiz Project [22],[23],[24],[25] of the IWiLL learning platform [54]. It provides English language learners online English reading materials collected from up-to-date online news websites and multiple-choice tests and automatically generates related quiz material. Here, English is used as the learning materials because the previous work, lexical knowledge acquisition [47], grammar [46], and lexical and semantic knowledge acquisition [45], have been shown a strong relationship between the age of acquisition and language learning.

AutoQuiz system explores the ways of generating "personalized" multiple choice questions in three question types, including vocabulary, grammar and comprehension [22]. The difficulty levels of words by the age of word acquisition - the temporal process by which learners learn the meaning and usage of new words. Then, similarly, difficulty levels of grammar patterns is modeled using the grade level of the textbook in which they frequently appear, which can be understood as a surrogate of the age of grammar acquisition. On the other hand, the difficulty of the reading comprehension questions was based on the reading level of the reading materials themselves. In this work, the grade level of the vocabulary and grammar questions are defined according to the semesters of high school in which the correct answer is taught, while the difficulty of the reading comprehension questions are measured by a reading difficulty estimation [55]. In other words, the grade level in this work is defined from one to six, corresponding to the six semesters of senior high school.

For vocabulary questions, AutoQuiz system selects distractor candidates of the same difficulty, part-of-speech (POS), similar character length and small Le-venshtein distance. For grammar questions, AutoQuiz system first use the Stanford Parser [56] to produce constituent structure trees of sentences and then use Tregex [57] to extract the instances of the target grammar patterns in trees. For comprehension questions, AutoQuiz focuses on the relation between sentences to generate two kinds of meaningful reading questions based on noun phrase co-reference resolution. The purpose of noun phrase co-reference resolution is to determine whether two expressions refer to the same entity in real life. The system first preprocesses an article to get coreference information [58] and transforms it into true and false statements with the help from linguistic resources, such as name entity recognition. The details can be found in [22].

### 3.2 Definition

To applying for computer-aided question generation, this paper proposes a novel and statistical method of estimating ability with inherent randomness in the acquisition process. For example, if a student is estimated as a certain grade level (i.e. grade level six), our method is able to estimate because the student has acquired a certain percentage of the knowledge in a population, when he correctly answers a certain percentage of items in a test. At first, we propose the following interpretation of the quantitative definition: an examinee is said to have ability $\theta$ if $s$ percent of items in a test $T = (t_1, \ldots, t_m)$ have been correctly answered each by $r$ percent of the population. In this example, when $s$ is denoted as 90% and $r$ is denoted as 80%. This student is estimated as a certain level because he answered correctly 90 percent of items in a test and this behavior is equal to 80 percent of the population.

### 3.3 The current ability estimation

We first consider that each item $t_i$ in a test $T$ has been correctly answered by $r$ percent of the population. In general, there is a specific knowledge behind each tested item $t_i$. The difficulty level of the specific knowledge represents the age at which most people have acquired knowledge of $t_i$. Most people understand some knowledge at an early age, whereas some understand this knowledge later in life. Here, we precisely denote the level the specific knowledge represents as the age at which $r$ percent of the population has acquired knowledge of $t_i$, where age can refer to school grades or lifetime. When given a knowledge $t_i$ and a population, the probability distribution of knowledge acquisition $p_t(\theta)$ can be calculated. Let the quantile function $q_t$ of the cumulative distribution function correspond to the acquisition distribution $p_t$. In other words, $q_t(r)$ represents the age $\theta$ at which $r$ percent of the population has acquired knowledge of $t$. This assumes a normal distribution,

$$q_t(r) = \mu_t + \Phi^{-1}(r)\sigma_t \qquad (1)$$

where $\mu_t$ and $\sigma_t$ represent the mean and standard deviation of the distribution $p_t$, and $\Phi^{-1}(r)$ is a quantile function representing the probability of exactly $r$ to fall inside the interval of the distribution. When an examinee correctly responds to the item $t_i$, the examinee's ability $\theta$ is regarded as the age or grade level, etc.

In practice, this is time consuming and costly to find the distribution $p_t$ for each item $t_i$ known in advance. Fortunately, under Item Response Theory, a response of an examinee to an item is modeled by a mathematical item response function, known as the item characteristic curve. The item characteristic curve is a mathematical family model that describes the probability of a correct response between an examinee's ability and the item parameters. These models employ one or more parameters, such as an item difficulty parameter or an item discrimination parameter, to define a particular cumulative form. When given the item parameters, the grade level at which $r$ percent of the population correctly responds to item $t$ can be inferred. Take one-parameter logistic model as an example,

$$q_t(r) = \ln(r/1-r) + b \tag{2}$$

where variable $b$ as item difficulty.

Estimating an examinee's current ability through a test relies on the test responses of the test. We model $s$ percent of items in a test which is correctly answered and aggregate the percentage of correct responses in the test. Similar to the idea from Classical Test Theory, where the item difficulty is the proportion of examinees who answer an item correctly in a sample, we investigate the distribution of the grade level of a test $T$. We collect the grade level values generated from each quantile function $q_t(r)$ as the distribution of knowledge acquisition within a single test $f_Q$. And then we consider a percentage of correct responses in a test as variable $s$ and find the $s$th quantile of the distribution of knowledge acquisition in a test $f_Q$ as the examinee's ability. The distribution of the $s$th quantile of $f_Q$, where s percent of items in a test have been correctly answered by $r$ percent of the population, can be performed using a standard formula for normal approximation of order statistics [59]:

$$q_T(r,s) \sim N(F_Q^{-1}(s), s(1-s)/m[f_Q(F_Q^{-1}(s))]^2) \tag{3}$$

where $F_Q$ is the cumulative distribution function and $m$ is the number of items in a test. This result is more certain of the estimated grade level assigned to a large sample item size. In cases where an examinee correctly answered all items or no item, a smooth constant $c$ is used ($c$=0.01 in this study).

### 3.4 The ability estimation with historical data

When given an examinee's responses in a test, the current examinee's ability $\theta_t$ can be described by the distribution (3) in which $r$ percent of the population correctly answer $s$ percent of items. We also consider an examinee's history record, and employ Exponential Moving Average (EMA) [60] to combine this history with the current ability, transformed by the following formula:

$$ability_t = \alpha \times \theta_t + (1-\alpha) \times ability_{t-1} \tag{4}$$

where $\theta_t$ is the current ability in time $t$ obtained from the mean of (3), $ability_{t-1}$ is the past estimated ability in the time $t-1$ as history records, and ability $t$ is the final estimated ability in time $t$ after the combination of the current ability and the past estimated ability with EMA. Additionally, $\alpha=2/(n+1)$ is a smoothing constant represented as an exponential weight, and $n$ represents the period as the length of the moving window.

### 3.5 Summary

When a teacher consider items in a test $T$ should be correctly answered each by $r$ percent of the population and an examinee correctly answered $s$ percent of the items in a test $T$, the procedure of estimating ability of the examinee can be summarized as the following. At first, each item in a test having been correctly answered by $r$ percent of the population is considered. Generally, there is a specific knowledge behind each tested item. The level of the specific knowledge represents that the age when the most of people have acquired the knowledge. The probability distribution of knowledge acquisition can be calculated in (1) when a knowledge and a population are given. When this information is incomplete and unavailable, this is impractical to know the distribution of knowledge behind each item in advance. To address this, we can employ (2) from Item Response Theory to calculate the percentage of people has correctly answered in an item. Moreover, we need to aggregate $s$ percentage of correct responses in a test. We address this by formulating the quantile of the distribution of acquisition within the test which is performed using (3) with the normal approximation of order statistics. Finally, (4) with Exponential Moving Average is employed to combine this history with the current ability.

## 4 SIMULATION

In this section, the proposed ability estimation is evaluated by a simulation study. To investigate the performance and the characteristics of the proposed method, we first analyze the convergence speed and the error distance between the ground truth and the estimated ability, and compared the proposed ability estimation with other related work. Next, an example, which presents the benefits of taking historical data into consideration, is shown. Finally, we evaluated the proposed method when the learning factor was considered. The details of the experimental designs are described in the following subsections.

### 4.1 Simulation method

To understand the performance of the proposed method, we conducted a simulation. Fig. 1 presents the procedure of the simulation.

Step 1: At first, we generated a simulated ability from a random sample of a population. Ability and difficulty in this study range from one to six, corresponding to the school grades. In practice, an examinee's school grade is considered as their initial ability (with standard deviation 0.2), for example, the simulated student is the fifth grade. We sample the initial ability from N(5, 0.2). And then the estimated ability is updated by responses in each test.

Step 2: In each simulation, ten items were generated according to an examinee's ability at the time. The distribution of difficulty of these items acts as a normal distribution. For example, given an examinee's ability $\theta=3$ and the number of items is 10, the difficulties of a test are {2, 2, 3, 3, 3, 3, 3, 3, 4, 4}.

Step 3: According to a one-parameter logistic model in Item Response Theory, the probability of correct response is 0.5 when an item difficulty is equal to an examinee's ability. In the simulation, we referred to this probability for setting the variable $r$. Moreover, the item response model also provides information in the estimation of the variable $s$. We used the one-parameter logistic model to predict the probability of a correct response when given the ability (the ground truth) and an item.

Step 4: We found the convergence point and then counted the Root Mean Square Error (RMSE) during the 100 iterations. The definition of the convergence point is determined by computing the difference between the estimated ability and the ground truth, and the difference value is continuously four times smaller than a threshold ($thd$=0.25 in the simulation). RMSE is defined as:

$$RMSE = \sqrt{\sum_i (\theta_i - \hat{\theta}_i)^2 / k} \quad (5)$$

where $\theta$ is the actual ability as the ground truth, $\hat{\theta}$ is the estimated ability, $k$ is the number of the iterations. Here, $k$=100. This metric represents the average distance between the ground truth and the generated results. The smaller RMSE value indicates that the estimated ability is close to the ground truth.

Step 5: Each simulation from the step 1 to step 4 was processed 1000 times. The simulation starts with any grade ranging from one to six in order to simulate different grade students with various abilities. In addition, we also discuss the parameter $\alpha$ in (4). The parameter is presented in terms of $n$ time periods and represents the weight of the observation at the present time. The variable $n$ was set from one to twelve.

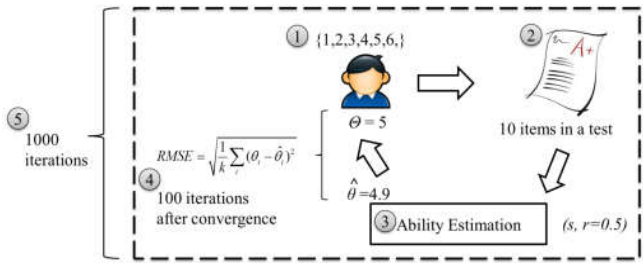

Fig. 1. The procedure of the simulation

### 4.2 Results

#### 4.2.1 Measurement precision

Table 1 shows the average convergence points in the number of variable $n$ of parameter $\alpha$ in (4) over the degree of difference between the estimated ability and ground truth, and the results of RMSE during the 100 iterations after the convergence points. It is clear that the proposed method can successfully estimate abilities in the finite iterations. Specifically, an examinee's ability can be estimated more precisely when he or she continues to have more tests. Furthermore, the error distances between the estimated abilities and the ground truths are low enough to be acceptable after convergence. That is, an examinee's ability can be steadily measured during a long-term observation.

The parameter $\alpha=2/(n+1)$ in the (4) is an exponential weight of the current ability, and $n$ represents the number of time periods, such as times or days, taken into consideration. When $n=1$, it represents that an examinee's ability only considers the current estimated ability without the history record. In Table 1, the values in screentone present that the average convergence points are fewer than the points generated from $n=1$. This result shows that the estimated abilities are quickly found and the error distances decrease when considering the history record. In particular, it is apparent when the initial grade is equal to the ground truth. When $n$ is small (e.g. $n=2$, $\alpha$ =2/3; $n=3$, $\alpha$ =1/2), the estimated ability is mainly decided by the current ability. The convergence points are smallest and the RMSE is slightly smaller than one generated from $n=1$. In contrast, when $n$ increases, the estimated ability is principally composed of abilities from the past to now. If an examinee's initial ability is not close to his or her actual ability, it takes more information to accurately estimate. Although it takes time, the RMSE is clearly shrinking.

#### 4.2.2 The comparison with other ability estimations

To understand the performance of the proposed ability estimation, we compare our results ($n=1$ used in this section) to those of MLE [3] and Lee [33]. One of the typical ability estimations in Item Response Theory is MLE in which the estimated ability is obtained by multiplying the item response function of each item and finding the highest possibility of which is the maximum likelihood estimate of a student's ability by using the Newton-Raphson method. Lee [33] extended BME in Item Response Theory and proposed a conventional approach to approximate the posterior distribution of the student's ability obtained from the subsequent responses.

Table 2 shows the results of RMSE between the proposed estimation and other estimations. Each row represents the degree of simulated student ability, and each column represents the given difficulty of a test. When the difficulty levels of items were equal to the abilities of simulated students (shown in the diagonals of the matrixes), the results estimated between MLE [3] and Lee [33] were similar, but those estimated by the proposed method were closer to the ground truth. With the increase in difference between the student abilities and item difficul-

TABLE 1
THE RESULTS OF CONVERGENCE POINT AND RMSE.

| d \ n | 1 | 2 | 3 | 4 | 5 | 6 | 7 | 8 | 9 | 10 | 11 | 12 |
|---|---|---|---|---|---|---|---|---|---|---|---|---|
| 0 | 20.61 | 13.88 | 11.72 | 11.53 | 10.98 | 10.90 | 10.26 | 10.52 | 10.16 | 10.35 | 10.18 | 10.04 |
| 1 | 21.96 | 16.17 | 15.74 | 16.31 | 17.40 | 19.07 | 20.43 | 22.29 | 23.98 | 25.45 | 26.92 | 28.42 |
| 2 | 22.91 | 18.08 | 18.54 | 19.91 | 21.90 | 24.18 | 26.64 | 29.06 | 31.50 | 33.53 | 35.62 | 38.58 |
| 3 | 23.86 | 19.67 | 19.91 | 21.91 | 24.59 | 27.62 | 30.33 | 32.90 | 35.74 | 38.43 | 41.52 | 44.13 |
| 4 | 24.30 | 20.73 | 21.52 | 23.51 | 26.71 | 29.68 | 32.96 | 36.00 | 40.19 | 42.83 | 45.45 | 48.65 |
| 5 | 24.50 | 21.41 | 22.66 | 25.22 | 29.10 | 31.92 | 35.97 | 38.22 | 42.62 | 46.40 | 49.18 | 53.12 |
| RMSE | 0.39 | 0.32 | 0.28 | 0.26 | 0.24 | 0.23 | 0.22 | 0.21 | 0.20 | 0.19 | 0.19 | 0.18 |

*Each row represents the degree of difference between the initial ability and the actual ability, and each column represents the number of time periods considered by the exponential weight of the current ability.*

TABLE 2
THE RESULTS OF RMSE BETWEEN MLE [3], LEE [33] AND THE PROPOSED ABILITY ESTIMATION

| s \ t | 1 | 2 | 3 | 4 | 5 | 6 | s \ t | 1 | 2 | 3 | 4 | 5 | 6 | s \ t | 1 | 2 | 3 | 4 | 5 | 6 |
|---|---|---|---|---|---|---|---|---|---|---|---|---|---|---|---|---|---|---|---|---|
| 1 | 0.22 | 1.00 | 2.01 | 2.99 | 4.04 | 5.13 | 1 | 0.21 | 1.01 | 2.04 | 3.05 | 4.13 | 5.17 | 1 | 0.13 | 0.52 | 1.04 | 1.51 | 1.95 | 2.18 |
| 2 | 1.00 | 0.23 | 1.00 | 2.02 | 3.03 | 4.04 | 2 | 1.01 | 0.22 | 1.01 | 2.05 | 3.11 | 4.15 | 2 | 0.51 | 0.13 | 0.52 | 1.04 | 1.54 | 1.95 |
| 3 | 2.00 | 0.99 | 0.22 | 1.01 | 2.01 | 3.03 | 3 | 2.03 | 1.00 | 0.21 | 1.02 | 2.04 | 3.11 | 3 | 1.03 | 0.52 | 0.13 | 0.53 | 1.03 | 1.53 |
| 4 | 2.96 | 1.99 | 1.00 | 0.23 | 1.03 | 2.01 | 4 | 3.05 | 2.02 | 1.01 | 0.22 | 1.04 | 2.05 | 4 | 1.50 | 1.02 | 0.51 | 0.13 | 0.53 | 1.03 |
| 5 | 3.98 | 3.01 | 1.98 | 1.00 | 0.24 | 1.01 | 5 | 4.09 | 3.07 | 2.01 | 1.01 | 0.23 | 1.02 | 5 | 1.93 | 1.53 | 1.01 | 0.52 | 0.13 | 0.52 |
| 6 | 4.91 | 3.93 | 2.98 | 2.00 | 1.00 | 0.23 | 6 | 4.74 | 3.78 | 2.84 | 1.87 | 0.89 | 0.11 | 6 | 2.16 | 1.92 | 1.51 | 1.03 | 0.52 | 0.13 |
| | MLE [3] | | | | | | | Lee [33] | | | | | | | The proposed method | | | | | |

ties, it was obvious that the proposed estimation produced more accurate estimated abilities than the other estimations. When questions were more difficult (the upper-right of the matrixes) or easier (the bottom-left of the matrixes) than the abilities of students, all of these methods failed to estimate the correct student abilities because the uncertainty among responses was unpredictable. But the error ranges of the proposed method were mostly within two grades; by comparison, the error ranges of the MLE [3] and Lee's [33] method were from four to five grades. This demonstrates that the proposed method is robust, especially when a student's ability is unknown. Moreover, note that the proposed method used in this section did not incorporate historical data during the estimation. This means that the estimated abilities will be obtained more accurately if both of the current responses and the past performance are used in the ability estimation, as demonstrated in the previous section.

### 4.2.3 The characteristics of the proposed ability estimation

Consider a dramatic example to explain the properties of the proposed method. Assume that a first grade student, whose real ability is the sixth grade, learns and has a test in a web-based learning system once a day. Fig. 2 illustrates the changes in the estimated ability computed from the proposed method in different weights. The black horizontal line at the sixth grade represents the student's actual ability as the ground truth. The other curves depict the estimated abilities under the different weights: a red dotted line, $n=1$; a green solid line, $n=3$; a purple solid line, $n=6$; and a blue solid line, $n=12$. The mark labels on each line are the convergence points (the value is continuously four times smaller than $thd = 0.25$). It is clear that the estimated abilities are converging as $n$ decreases in size. Although these estimated abilities are estimated using few iterations when $n=1$, the red-dotted line drastically fluctuates after the convergence point. In other words, if the ability estimation only takes the current responses into consideration, instead of past performance, the variance of every estimated ability may be large. In this situation, question selection in a test using inaccurate ability estimation could result in confusion by the examinee. In contrast, the estimated error gradually decreases when $n>1$, even though the estimated abilities when $n=1$ take more time to estimate. In this situation, the students' abilities were gradually updated and the difficulties of items incrementally increased. This is thus a trade-off problem between speed and precision.

### 4.2.4 Estimating with the continuous improvement.

Learning is the temporal process of acquiring knowledge or skill. A learner's ability will change when time goes by. Glickman [61], Klinkenberg et al. [15] and Wang et al. [34] argued that the impact of time may lead to inaccuracies on estimation. Thus, Klinkenberge et al. [15] designed a $K$ factor to capture the uncertainty in their ability estimation.

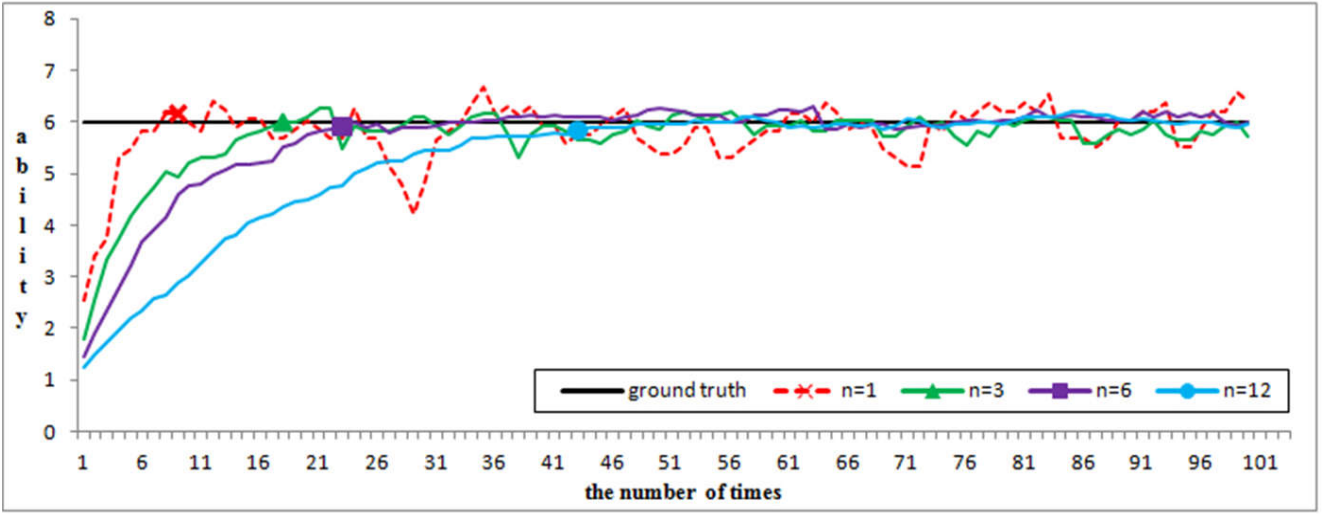

Fig. 2. The changes in the estimated ability computed from the proposed method for the different weights (n=1, n=3, n=6, n=12).

To understand the influence of the learning factor on ability estimation, we used a learning factor $l$ to reflect the uncertainty in the following experiment. Here, the learning factors were constants. Generally, a learner gains new knowledge every day and upgrades after a year. In the experiment, we assumed a student can upgrade from grade level one to two in a half year. The learning factor in a day can be denoted as $l_{normal} = 1/182 = 0.0054945$, and we sampled it from a normal distribution $l_{normal}$~N (0.0054945, 0.001). If a learner is good at the skill, the learning factor will be high, $l_{fast}$~ N (0.010989, 0.001). By contrast, if a learner have some difficulty in learning (i.e. a student takes a year to upgrade from grade level one to two), the learning factor will be low, $l_{slow}$~N (0.0027425, 0.001).

In this experiment, the learning factors would be added to the ground truth in each iteration in order to simulate the temporal process how a learner acquires knowledge. In each iteration, we conditionally sampled the learning factor in a normal distribution with variance 0.001, ie. $l_{normal}$~ (0.0054945, 0.001). We compared four conditions, $l_{normal}$, $l_{fast}$, $l_{slow}$ and $l_{no}$. The condition $l_{no}$ represents a student with a fixed ability as Section 3.2.1. The other setting in this comparison was the same in Section 3.1. Because we focused on the learning factors, we only reported the results when the initial grade and the initial ground truth were set as the same.

Fig. 3 reports the average convergence points in (a) and the RMSE during the 100 iterations after the convergence points in (b) among the different periods considered in the moving window. It is obvious that the proposed method still worked effectively even when estimating ability with the continuous improvement. Overall, the average convergence points were less than the result which the ground truth did not including the learning factor ($l_{no}$). Especially, the ability estimation took fewer iterations to predict when the learning factor was high and more historical data was considered. In terms of the error distance between the estimated abilities and the ground truths, the values increased when the learning factors were getting higher. The distance from the normal and slow learning factors were lower than 0.5 but the result from the fast learning factor was close to 0.7. During a long-term observation, the distance would gradually decrease.

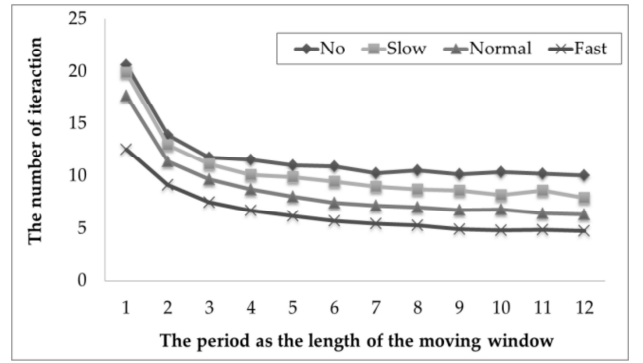

(a)

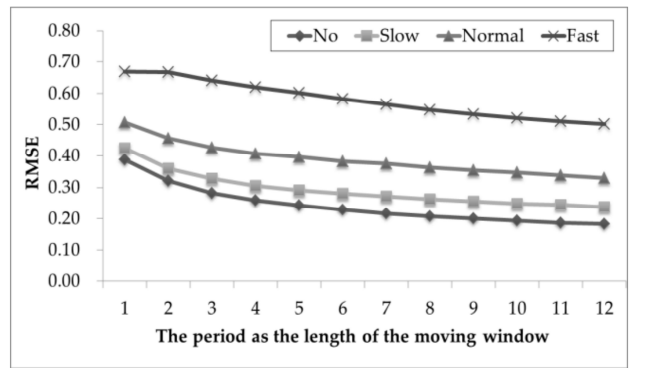

(b)

Fig. 3. The average convergence points in (a) and the RMSE values in (b) among the different periods considered in the moving window.

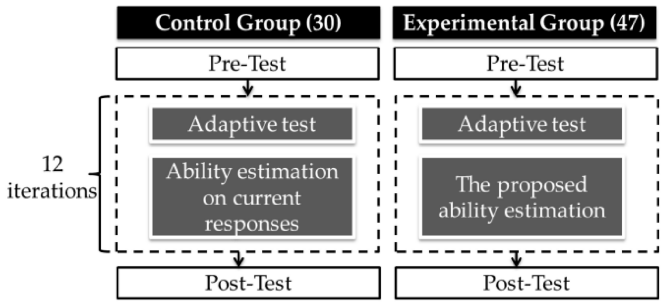

Fig. 4. The experimental procedure of the empirical study.

## 5 EMPIRICAL STUDY

In this section, the proposed ability estimation was examined by an empirical study. To investigate the performance of the proposed method with the empirical data, we first examined the the estimated abilities and real data. In addition, the students' performance was analyzed whether or not appropriate instructional scaffolding could help students advance their learning when effectively identifying their abilities. The details of the experimental designs are described in the following subsections.

### 5.1 Participants and procedure

The participants in this study were high school students in Taiwan, divided into two groups: a control group where ability is estimated only based on current responses, and an experimental group that incorporates the history record into the current ability estimation. 30 students participated within the control group, while 47 students participated in the experimental group.

The experiment was held from January 30th to March 4th, 2012. As Fig. 4 shown, during the experiment, the subjects were asked to participate in twelve activities, consisting of reading an article and then taking a test. In each activity, the subjects in both groups received an up-to-date article and a series of quizzes automatically generated based on their abilities. Each test was composed of ten vocabulary questions, five grammar questions, and three reading comprehension questions. A total of 2,425 items were automatically generated based on 72 reading materials. In addition, there was a pre-test and post-test for evaluating their abilities as the ground truth. The variable $r$ was set as 0.5 based on Item Response Theory, and the variable $s$ defined as the percentage of correctly answered items. Furthermore, the parameter $n=12$ in the exponential weight of the experimental group was equal to the period of activity, because all test records were taken into consideration.

### 5.2 Results

#### 5.2.1 Measurement validity

To validate the accuracy of the proposed ability estimation, the subjects' abilities in the two groups were estimated with twelve continuous activities. Table 3 reports the Pearson's correlation coefficient between the estimated abilities (the estimated grade is rounded by the esti-mated score) and the post-test scores among the three quiz types. All of the measures are significantly positively correlated. The results in the experimental group ranged from 0.44 to 0.69, while ones in the control group ranged from 0.47 to 0.54. Most of the correlation values in the experimental group are higher than the values in the control group; this suggests that estimating ability with the history record leads to a clearer relationship between the estimated ability and the ground truth.

TABLE 3
THE CORRELATION RESULT BETWEEN THE ESTIMATED ABILITY AND THE POST-TEST IN THE CONTROL GROUP AND THE EXPERIMENTAL GROUP.

|  | vocabulary | | grammar | | reading comprehension | |
|---|---|---|---|---|---|---|
|  | score | grade | score | grade | score | grade |
| **Control group** | 0.47* | 0.49** | 0.54** | 0.51** | 0.54** | 0.47* |
| **Experimental group** | 0.51*** | 0.44** | 0.55*** | 0.55*** | 0.69*** | 0.65*** |

*$p<0.05$, **$p<0.01$, ***$p<0.001$

Comparing the post-test score in each estimated ability (grade) is another way to assess the accuracy of the proposed ability estimation. If the estimated abilities are accurate, the subject performance of each ability will differ from that of other abilities. Table 4 presents the mean post-test score of the subjects of different estimated abilities between the control group and the experimental group. Intuitively, a subject estimated a higher ability should have higher post-test score than one estimated a lower ability. One-way Analysis of Variance revealed that there were differences in the estimated vocabulary ability (F=5.75, p=0.001), the estimated grammar ability (F=4.71, p=0.003) and the estimated reading comprehension ability (F=5.98, p<0.001) in the experimental group, while there were no statistical differences between the estimated vocabulary and grammar ability in the control group. Noticeably, although the estimated reading comprehension ability in the control group has a significant difference, the mean scores among every ability fluctuated. The bolded values in Table 4 are unreasonable, because the averaged scores of the higher estimated abilities (grade 2, grade 4 and grade 5) in the control group were lower than ones of the lower estimated abilities (grade 1 and grade 3). Though there was an unreasonable value for grade 6 of the estimated vocabulary ability in the experimental group, this is likely because only two students were assigned to grade 6. This sample size is likely unrepresentative. Moreover, in the experimental group, a Bonferroni post hoc test indicated that the performance of the estimated ability 1 and 2 were significantly different from the estimated ability 5 and 6. This indicates that the proposed ability estimation can effectively distinguish higher ability examinees from lower ones.

To evaluate the validity of the proposed ability estimation, a logistic regression was performed. Table 5 shows the equations using the ability of a student $i$ and the difficulty of a question $j$ on the log odds ratio of the observation, in which student $i$ correctly answering question $j$ is

TABLE 4
THE MEAN POST-TEST SCORE OF THE SUBJECTS IN DIFFERENT ESTIMATED ABILITY GROUPS BETWEEN BOTH GROUPS AND THE RESULT OF ANOVA.

| Esti-mated ability | Control group | | | Experimental group | | |
|---|---|---|---|---|---|---|
| | vocabu-lary | gram mar | read-ing | vocab-ulary | gram mar | read-ing |
| 1 | - | 37.50 | 46.80 | - | - | 37.67 |
| 2 | 48.33 | 47.00 | **40.00** | 23.00 | 34.33 | 46.63 |
| 3 | **38.00** | 51.40 | 52.57 | 52.86 | 52.80 | 53.50 |
| 4 | 54.40 | 41.40 | **41.00** | 62.33 | 54.94 | 64.50 |
| 5 | 61.22 | 62.83 | **32.67** | 69.71 | 66.81 | 66.90 |
| 6 | 65.83 | 65.56 | 70.18 | **57.67** | 72.00 | 78.00 |
| F score | 2.67 | 2.54 | 6.12*** | 5.75*** | 4.71** | 5.98*** |

***p<0.01, ***p<0.001*

in class 1 or the student incorrectly answering question $j$ is in class 0. Generally, the probability of a question being correctly answered is relatively higher when the ability of a student is more advanced. On the other hand, the more difficult a question is, the lower the probability of a student correctly answering the question. If the observed abilities in the empirical study are precisely estimated, the relationship between the estimated abilities and dichotomous outcome will be explainable. The results indicates that the regression coefficients for the ability of each student among these three question types are positive and the coefficient values for the difficulty of each question among these types are negative. Even though the values among three question types are slightly different, all of them had the same influence on the dependent variable. This supports the assumption that the estimated abilities of students were so accurate that students of advanced proficiencies could correctly respond to more difficult questions.

TABLE 5
THE EQUATIONS AMONG QUESTION TYPES REPRESENT THE LOG ODDS RATIO OF THE OBSERVATION THAT STUDENT I CORRECTLY ANSWERING ITEM J IS IN CLASS 1 OR THE STUDENT INCORRECTLY ANSWERING ITEM J IS IN CLASS 0.

| Question types | Equations |
|---|---|
| vocabulary | $ln(p_{ij}/1-p_{ij})=-1.554+1.129student_i-0.321question_j$ |
| grammar | $ln(p_{ij}/1-p_{ij})=-1.518+0.859student_i-1.321question_j$ |
| reading comprehension | $ln(p_{ij}/1-p_{ij})=-0.178+0.898student_i-0.783question_j$ |

### 5.2.2 The learning performance of students

To further understand the impact of employing the proposed ability estimation on learners, we investigated the performance between the control group and the experimental group. In keeping with the previous results, the estimated subjects' abilities in the experimental group were more accurate than those in the control group. We assume that appropriate instructional scaffolding could help students advance their learning, when effectively identifying their abilities. Table 6 presents the descriptive statistic and results of a T-test between the pretest and post-test. The results of the independent T-test (p=0.92 in the pre-test and p=0.51 in the post-test) showed a similar effect on the post-test between the experimental group and the control group. One explanation for the results may be rooted in the short time (only five weeks) allowed for the treatment in the experiment, while Klinkenberg et al. [15] conducted one-year experiment and Barla et al. [8] employed their method for a winter term course. However, it is noticeable that the average score of the experimental group in the pretest was lower than the control group, but that of the experimental group in the post-test made great progress and surpassed the control group. Additionally, the paired sample T-test showed a significant effect of the pre-test and the post-test in the experimental group (p<0.001), while the performance of the control group had no statistically significant effect (p>0.05). This indicates that the subjects in the experimental group with an appropriate support can exceed the past themselves when successfully recognizing their learning status.

TABLE 6
THE RESULTS OF THE PRETEST AND POST-TEST BETWEEN THE CONTROL GROUP AND THE EXPERIMENTAL GROUP.

| | Pretest | | Post-test | | Paired sample |
|---|---|---|---|---|---|
| | mean | std. | mean | std. | t-test |
| Control group | 53.23 | 19.35 | 56.70 | 17.99 | 1.57 |
| Experimental group | 52.83 | 16.67 | 59.28 | 16.01 | 3.71*** |
| independent t-test | 0.20 | | 0.66 | | |

*\*\*\*p<0.001*

## 6 CONCLUSION

This work develops an alternative method of estimating ability that captures the succession of learning over time in a personalized computer-aided question generation testing environment. Moreover, it brings a greater flexible measurement based on the quantiles of acquisition distributions instead of pre-calibration. This method draws a connection between students' abilities and the acquisition age distribution. The results from the simulation demonstrate that the estimated abilities obtained from the proposed method could successfully approximate the simulated abilities of students, and estimated abilities can be steadily measured during long-term observation. Compared with other ability estimations, the proposed ability estimation gives the lowest root mean square error among all of the estimations. And more specifically noteworthy is that with an increase of distance between a student's ability and an item difficulty, the proposed method yields the most accurate results. This shows that the proposed estimation is the most robust when the responses were uncertain, i.e., for questions with degrees of difficulty less than or greater than the abilities of students. This proposed approach was also implemented on a computer-aided question generation. The empirical results reveal that the correlation values incorporating this testing history were higher than the values that only consider the test responses at the time of testing. Students who were estimated as advanced graders showed significantly higher post-test scores and better responses than ones who were estimated as basic graders. Additionally, the pretest and post-test administered to the experimental group demonstrated significant student improvement. These results suggest that this method can serve as a successful alternative ability estimation and can provide a better understanding of student competence.

To the best of our knowledge, there has been no prior work on our research topic, which is to estimate an examinee's ability based on the distribution of the age of acquisition. Through this idea, for example, the estimated ability represents a student as grade level six because the examinee answered correctly 90 percent of the items in a test with the difficulty level normally distributed at level six, and this behavior is equal to 80 percent of the population (assume $s$=90% and $r$=80%). Unlike the traditional approaches, which focus on a norm referenced item parameter scale for an individual item, the ability parameter estimated by the proposed method is explainable in terms of an ability scale that is based on the age that most people acquire the certain piece of knowledge. In addition, the proposed method estimates an examinee's ability from all the responses of questions in a test; in contrast, the traditional approaches determine the ability by an individual question. This point is similar to Classical Test Theory [3],[4],[5], which considers all responses in a test as an examinee's observed scores. However, while the results of Classical Test Theory are sample-dependent, the estimated result from our proposed method is stable due to the estimation being based on the acquisition age distribution. Moreover, our estimated ability is obtained from the weighted combination of an examinee's current performance and his or her historical data. The large amount of historical data enables more accurate ability estimates. This characteristic inherits the strength of BME [31] [32] [33], which consider the successive change in the ability level within a learning session, and achieve more accurate results than the BME.

The ability estimation described in this study enables students and teachers to see the estimated ability from different perspectives. First, it can be used as a guideline for identifying the current learning status of students for providing instructional supports, which could in turn indicate what students have not acquired yet. For example, when the estimated ability of a student is determined, the student could better understand his or her learning status because the ability is estimated based on the difficulty levels of words he or she acquired. It is thus easier for students to see the extent of their proficiency in the context of different levels. Also, it could be used in a quantitative purpose for adapting to different learning environments while still offering flexible measurement, such that different values could be set for the two parameters $r$ and $s$ depending on the various conditions. A good example is native speakers versus second language learners. In this way, teachers could adjust the parameters of the proposed ability estimation to the test purpose, regarding a qualified ability corresponding to the age which the certain percent of a population have acquired the piece of knowledge.

The proposed ability estimation with advanced technology affords the possibility of providing a better learning environment. For example, with the benefits of cloud computing, learner activities could be recorded on any platform as an e-portfolio [62]. A student's ability can be estimated by being integrated with the historical data in the e-portfolio, instead of only considering the current responses. Another example involves Big Data [63]. With the emergence of abundant online learning materials and electronic textbooks, it has become very feasible to sample the quantiles of acquisition distributions via the application of Natural Language Processing[51], Information Retrieval [49],[50] and statistically analytical methods [52] [53]. It could be practical for our proposed ability estimation to be implemented in any field.

One of limitations of the study is that the distribution of item difficulties of the questions in a test was assumed as a normal distribution. Even though teachers usually design a combination of difficulties of questions in a test which is similar to a normal distribution, some questions are uniformly generated. One possible solution is that the item discrimination parameter and the guessing parameter described in three-parameter logistic model of Item Response Theory might be taken into consideration. The item characteristic curve could accurately model the probability of a correct response between an examinee's ability and the item parameters. This important concern can be profitably taken into account in future research. Another limitation is that take account the incorrect responses. One of potential future direction is to find a way of combining this proposed method with penalizing the estimated value when an examinee selecting an incorrect answer. Furthermore, it is worth noting that the experimental materials used in the empirical study were in the domain of language. Although vocabulary, grammar, and reading comprehension are the basic skill of language learning, the nature of these skills are different from each other and the experimental results showed the estimated abilities among these skills were positively consistent. Similarly, it would be important to replicate the proposed ability estimation with all kind of learning domains in the future.

In conclusion, the proposed method meets the requirements in the personalized learning environment by statistically interpreting the ability based on acquisition age distributions, and considers long-term observation as a student's estimated ability. It is the first to mathematically draw a connection between ability estimation and the age of acquisition, and to successfully evaluate simulated and empirical data for estimating the grade level of a student. Interestingly, from the results of the simulation and the empirical study, the proposed model performs very well in estimating the ability of a student in practice while also providing students learning with appropriate support. Further research will extend the proposed ability estimation to examine the impact of adaptive learning to another field. We look forward to a fast adoption of this kind of learning environment and hope students and teachers will be able to draw on the benefits of this work.

## ACKNOWLEDGMENT

This article combines material from [64] on the the preliminary results, and substantial new content including the evaluations and explainations. This work was partially supported by National Science Council, Taiwan, with Grant No. 100-2511-S-008-005-MY3 and NSC97-2221-E-001-014-MY3.